\documentclass[11pt,twoside]{article}
\usepackage{asp2010}

\resetcounters

\begin{document}

\title{A Tale Of 160 Scientists, Three Applications, A Workshop and A Cloud}
\author{G. Bruce Berriman$^1$, Carolyn Brinkworth$^1$, Dawn Gelino$^1$, Dennis K. Wittman$^1$, Ewa Deelman$^2$, Gideon Juve$^2$, Mats Rynge$^2$, Jamie Kinney$^3$
\affil{$^1$NASA Exoplanet Science Institute, California Institute of Technology, Pasadena CA 91125, USA}
\affil{$^2$ USC Information Sciences Institute, 4676 Admiralty Way, Marina del Rey, CA 90292, USA}
\affil{$^3$ Amazon Web Services, 1918 8th Ave, Seattle, WA 98101, USA}}

\begin{abstract}
The NASA Exoplanet Science Institute (NExScI) hosts the annual Sagan Workshops, thematic meetings aimed at introducing researchers to the latest tools and methodologies in exoplanet research. The theme of the Summer 2012 workshop, held from July 23 to July 27 at Caltech, was to explore the use of exoplanet light curves to study planetary system architectures and atmospheres. A major part of the workshop was to use hands-on sessions to instruct attendees in the use of three open source tools for the analysis of light curves, especially from the Kepler mission. Each hands-on session involved the 160 attendees using their laptops to follow step-by-step tutorials given by experts. 

One of the applications, PyKE, is a suite of Python tools designed to reduce and analyze Kepler light curves; these tools can be invoked from the Unix command line or a GUI in PyRAF. The Transit Analysis Package (TAP) uses Markov Chain Monte Carlo (MCMC) techniques to fit light curves under the Interactive Data Language (IDL) environment, and Transit Timing Variations (TTV) uses IDL tools and Java-based GUIs to confirm and detect exoplanets from timing variations in light curve fitting. 

Rather than attempt to run these diverse applications on the inevitable wide range of environments on attendees laptops, they were run instead on the Amazon Elastic Cloud 2 (EC2). The cloud offers features ideal for this type of short term need: computing and storage services are made available on demand for as long as needed, and a processing environment can be customized and replicated as needed. The cloud environment included an NFS file server virtual machine (VM), 20 client VM’s for use by attendees, and a VM to enable ftp downloads of the attendees' results. The file server was configured with a 1 TB Elastic Block Storage (EBS) volume (network-attached storage mounted as a device) containing the application software and attendees’ home directories. The clients were configured to mount the applications and home directories from the server via NFS. All VM’s were built with CentOS version 5.8. Attendees connected their laptops to one of the client VMs using the Virtual Network Computing (VNC) protocol, which enabled them to interact with a remote desktop GUI during the hands-on sessions. 

We will describe the mechanisms for handling security, failovers, and licensing of commercial software. In particular, IDL licenses were managed through a server at Caltech, connected to the IDL instances running on Amazon EC2 via a Secure Shell (ssh) tunnel. The system operated flawlessly during the workshop.

\end {abstract}

\section{Introduction}
The NASA Exoplanet Science Center (NExScI) hosts the Sagan Workshops, which are annual themed conferences aimed at introducing the latest techiques in exoplanet astronomy to young researchers. The workshops emphasize interaction with data, and include hands-on sessions where participants use their laptops to follow step-by-step tutorials given by experts.  The 2012 workshop had the theme "Working With Exoplanet Light Curves," and posed special challenges for the conference organizers because the  three applications chosen for the tutorials run on different platforms, and because over 160 persons attended,  the largest attendance to date. One of the applications, PyKE, is a suite of Python tools designed to reduce and analyze Kepler light curves; it is called from the command line or from a GUI in PyRAF. The Transit Analysis Package (TAP) uses Markov Chain Monte Carlo (MCMC) techniques to fit light curves in the Interactive Data Language (IDL) environment, and Systemic Console analyzes Transit Timing Variations (TTV) with IDL and Java-based GUIs to confirm and detect exoplanets from timing variations in light curve fitting.  Rather than attempt to run these diverse applications on the inevitable wide range of environments on attendees' laptops, the conference organizers, in consulation with the Virtual Astronomical Observatory, chose instead to run the applications on the Amazon Elastic Cloud 2 (EC2). This paper summarizes the system architecture, the Amazon resources consumed, and lessons learned and best practices.

\section{The System Architecture}

The Sagan Workshop took advantage of the EC2's capabilities to support Virtual Machines (VMs) that can be customized to meet local needs, then replicated, and then released on completion of the jobs. Fig. 1 shows the system architecture developed to support the Sagan Workshop. Participants logged into one of 20 tutorial servers via a Virtual Network Connection (VNC). The Amazon Elastic Block Storage (EBS) system and the Network File System (NFS) were used to share common datasets and user home directories across all virtual machines. An IDL license server at IPAC received license request through an ssh tunnel. The following list describes the architecture component by component and the rationale for the design choices.

\articlefigure {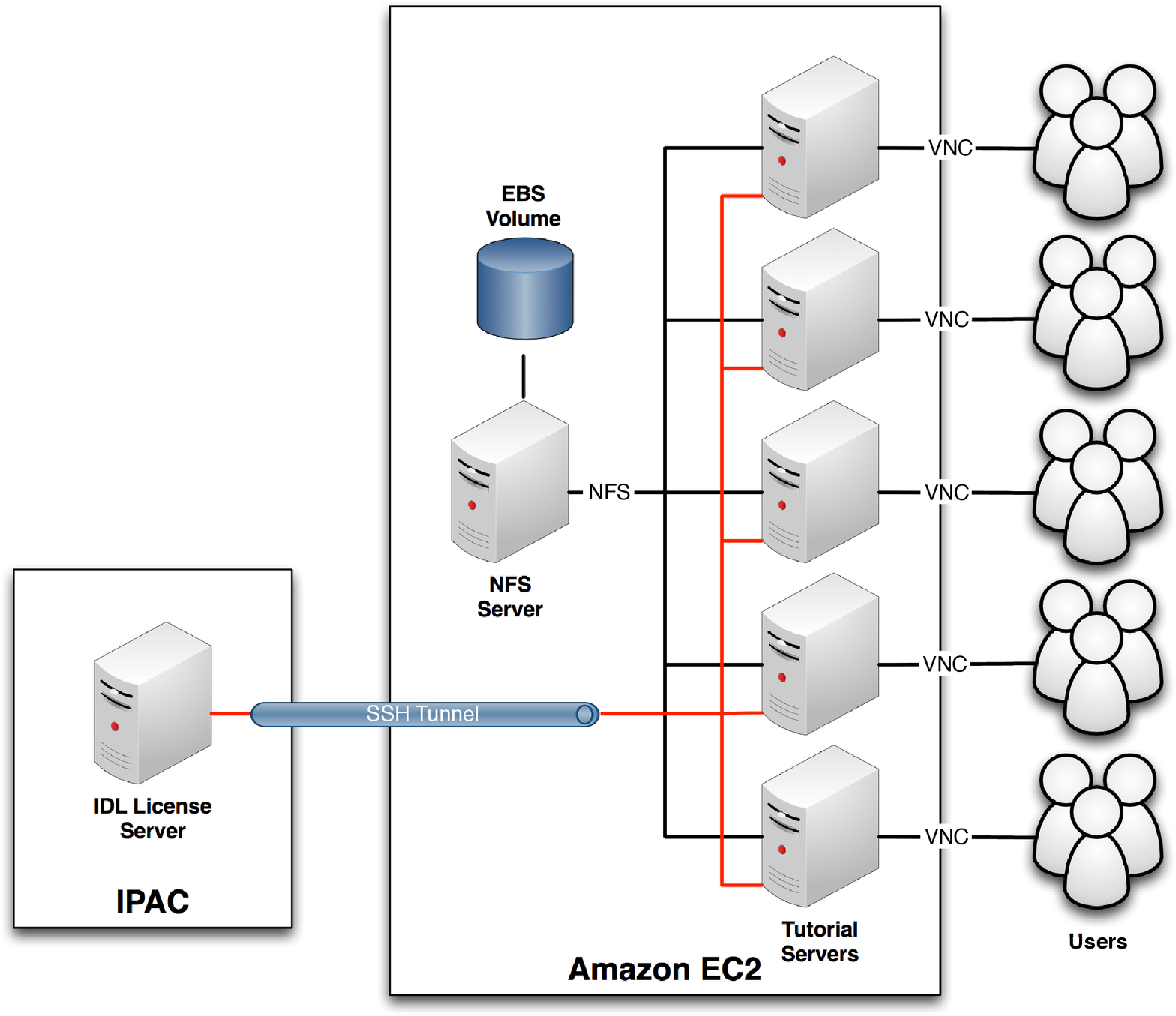}{Fig 1}{Hardware architecture to support the Sagan Workshop}

\begin{itemize}

\item One master virtual machine image, built on the Cent OS 64-bit operating system, was used for all servers.
A boot script determined the VM’s identity.
Usernames and passwords were the same on all machines. 
\item
1 TB of Elastic Block Storage (EBS), a 
block-based storage service where
volumes appear as disk drives connected to VMs,
contained applications, tutorial data, and user home directories. Applications and tutorial data are installed on VM images, and so data are not lost if a tutorial server fails.

\item
The EC2 m1.2xlarge instance type was chosen to handle the load of 20 tutorial servers. It has enough memory to cache commonly accessed files,  mounts all the partitions from the EBS volumes, and exports all partitions via NFS to the tutorial servers.
\item
The tutorial servers were EC2 c1.xlarge instance type, with 
8 cores and 7 GB RAM, chosen because the applications were CPU-bound.
Server performance was benchmarked with 8 users, but the servers were in fact able to support up to 25 users. 
\item
A Virtual Network Computing (VNC) server provided remote desktop logins to the tutorial servers. VNC is similar to the X Window system, but sends compressed images instead of drawing commands and proved more responsive than X in our tests. Each tutorial server ran one VNC server that supported up to 30 connections.  Screen resolution set to 1024x768 to balance usability and performance. In practice,  the workshop used TigerVNC as the server and RealVNC as the client.
\item
The tutorial servers were connected via an ssh tunnel to an IDL license server at IPAC. 
IDL VM sessions think the license server is on localhost, and the license server thinks IDL is inside IPAC's network. We used autossh to ensurethe tunnel was re-established if disconnected
\item
The Amazon AWS Security Rules limited access only to the VNC, SSH and IDL ports, and only from the Caltech and IPAC subnets  used to support the workshop.
\end{itemize}

\section{Cost of Using the Amazon Elastic Cloud 2}

Had the Sagan Workshop's Amazon EC2 costs not been met by an educational grant, the total cost of installation, testing and running the workshop sessions would have been \$2,876. The breakdown of the costs is shown in Table 1.

\begin{table}[!ht]
\caption{Breakdown Of The Costs of Using Amazon EC2 During The Sagan Workshop}
\smallskip
\begin{center}
{\small
\begin{tabular} {llr}
\tableline
\noalign{\smallskip}
Resource & Consumption & Cost (\$)\\
\noalign{\smallskip}
\tableline
\noalign{\smallskip}
VM Instances & 4,159 hours &  2,738 \\
EBS Storage & 1.25 TB & 126 \\
I/O Requests & 12 million & 1 \\
Snapshot data storage & 22 GB & 3 \\
Use of elastic IP addresses & 604 hours &  3 \\
Data Transfer & 55 GB & 5\\
Total & ... & 2,876 \\
\noalign{\smallskip}
\tableline
\end{tabular}
}
\end{center}
\end{table}

\section{Lessons Learned and Best Practices}
These may be summarized as follows:

\begin{itemize}
\item  Automate processes wherever possible, as this allows easier management of large numbers of machines and easy recovery in the case of failure. Tutorial servers automatically mounted NFS partitions when booted and SSH tunnels automatically reconnected on failure.
\item Test, test, and test again. Document and test all the steps required to recover if a VM fails, and step through the tutorials under as close to operational conditions as possible.
\item Develop a failover system. We copied the final software configuration to two local machines for use if Amazon failed.
\item Give yourself plenty of time to solve problems. In our case, we needed to assure the IDL vendor that licenses would not persist on the cloud, and we needed to understand the poor performance of X for remote access to the cloud.
\end{itemize}

\acknowledgements The Sagan Workshop was funded as part of the Sagan Program through NASA's Exoplanet Exploration Program. We thank Amazon Web Services for the award of a generous Educational Gran. ED, GJ and MR acknowledge support through NSF OCI-0943725. The VAO is jointly funded by NSF and NASA, and is being managed by the VAO, LLC, a non-profit 501(c)(3) organization registered in the District of Columbia and a collaborative effort of the Association of Universities for Research in Astronomy (AURA) and the Associated Universities, Inc. (AUI). We thank Dr. Peter Plavchan for suggesting we examine VNC.


\end{document}